\documentclass[aps, prd, 12pt, eqsecnum, tightenlines, notitlepage, nofootinbib, floatfix]{revtex4-1}

\PassOptionsToPackage{unicode}{hyperref}

\usepackage{hyperref, graphicx, slashed, xcolor, amsmath}

\allowdisplaybreaks

\definecolor{red}{rgb}{1,0,0}

\newcommand{\beq}{\begin{equation}}
\newcommand{\eeq}{\end{equation}}
\newcommand{\bea}{\begin{eqnarray}}
\newcommand{\eea}{\end{eqnarray}}

\begin{document}

\normalsize
\vspace{-0.5cm}
\begin{flushright}
CALT-TH-2017-071\\
\vspace{1cm}
\end{flushright}


\title{Loop-Induced Stochastic Bias at Small Wavevectors}

\author{Michael McAneny, Alexander K. Ridgway, Mikhail P. Solon and Mark B. Wise}
\affiliation{Walter Burke Institute for Theoretical Physics,
California Institute of Technology, Pasadena, CA 91125}

\begin{abstract}
Primordial non-Gaussianities enhanced at small wavevectors can induce a power spectrum of the galaxy overdensity that differs greatly from that of the matter overdensity at large length scales. In previous work, it was shown that ``squeezed" three-point and ``collapsed" four-point functions of the curvature perturbation $\zeta$ can generate these non-Gaussianities and give rise to so-called scale-dependent and stochastic bias in the galaxy overdensity power spectrum.  We explore a third way to generate non-Gaussianities enhanced at small wavevectors: the infrared behavior of quantum loop contributions to the four-point correlations of $\zeta$.  We show that these loop effects lead to stochastic bias, which can be observable in the context of quasi-single field inflation.
\end{abstract}

\maketitle

\section{Introduction}
The inflationary paradigm \cite{SKS} proposes an era in the very early universe during which the energy density is dominated by vacuum energy and the universe undergoes exponential expansion.  Such a period elegantly explains why the universe is close to flat and the near isotropy of the cosmic microwave background (CMB).  It also provides a simple quantum mechanical mechanism for generating energy density perturbations which have an almost scale-invariant Harrison-Zel'dovich power spectrum.

The simplest inflation models consist of a single scalar field $\phi$, called the inflaton, whose time-dependent vacuum expectation value drives the expansion of the universe.  The quantum fluctuations in the Goldstone mode $\pi$ associated with the breaking of time translation invariance by the inflaton~\cite{Cheung:2007st} 
source the energy density fluctuations. In the simplest of these single field models, the density perturbations are very nearly Gaussian~\cite{Maldacena:2002vr}.  One way to generate measurable non-Gaussianities is to introduce a second field $s$ that interacts with the inflaton field during the inflationary era. A simple realization of such a model is quasi-single field inflation (QSFI)~\cite{Chen:2009zp}.

These non-Gaussianities affect the correlation functions of biased tracers of the underlying matter distribution such as galaxies.  It was first pointed out in~\cite{AGW} and~\cite{Dalal:2007cu} that the power spectrum of the galaxy overdensity can become greatly enhanced relative to the Harrison-Zel'dovich spectrum on large scales if the primordial mass density perturbations are non-Gaussian.\footnote{We refer to these effects as ``enhancements" even though for certain model parameters they can interfere destructively with the usual Gaussian primordial density fluctuations.} These enhancements are known as scale-dependent bias and stochastic bias and were systematically explored in the context of QSFI in \cite{Baumann:2012bc} and \cite{An:2017rwo}.

The enhancements studied in~\cite{AGW} and~\cite{Dalal:2007cu} result from tree-level contributions to the three- and four-point functions of $\pi$ that are in their ``squeezed" and ``collapsed" limits.  In this paper, we consider quantum contributions to the correlation functions of $\pi$ which can also give rise to these long-distance effects.  We find that the infrared region of loop integrals can induce sizable stochastic bias on large scales without introducing any scale-dependent bias.  In section \ref{Loop Induced Stochastic Bias} we illustrate this loop effect using a higher dimension operator that would appear in a generic effective theory of multi-field inflation. In section \ref{qsfi section} we show that the loop effect can be observable in the context of QSFI and estimate the distance scale at which the loop contribution to the galaxy power spectrum could exceed the usual Harrison-Zel'dovich one.

\section{Loop-Induced Stochastic Bias}
\label{Loop Induced Stochastic Bias}
\begin{figure}
\includegraphics[width=3in]{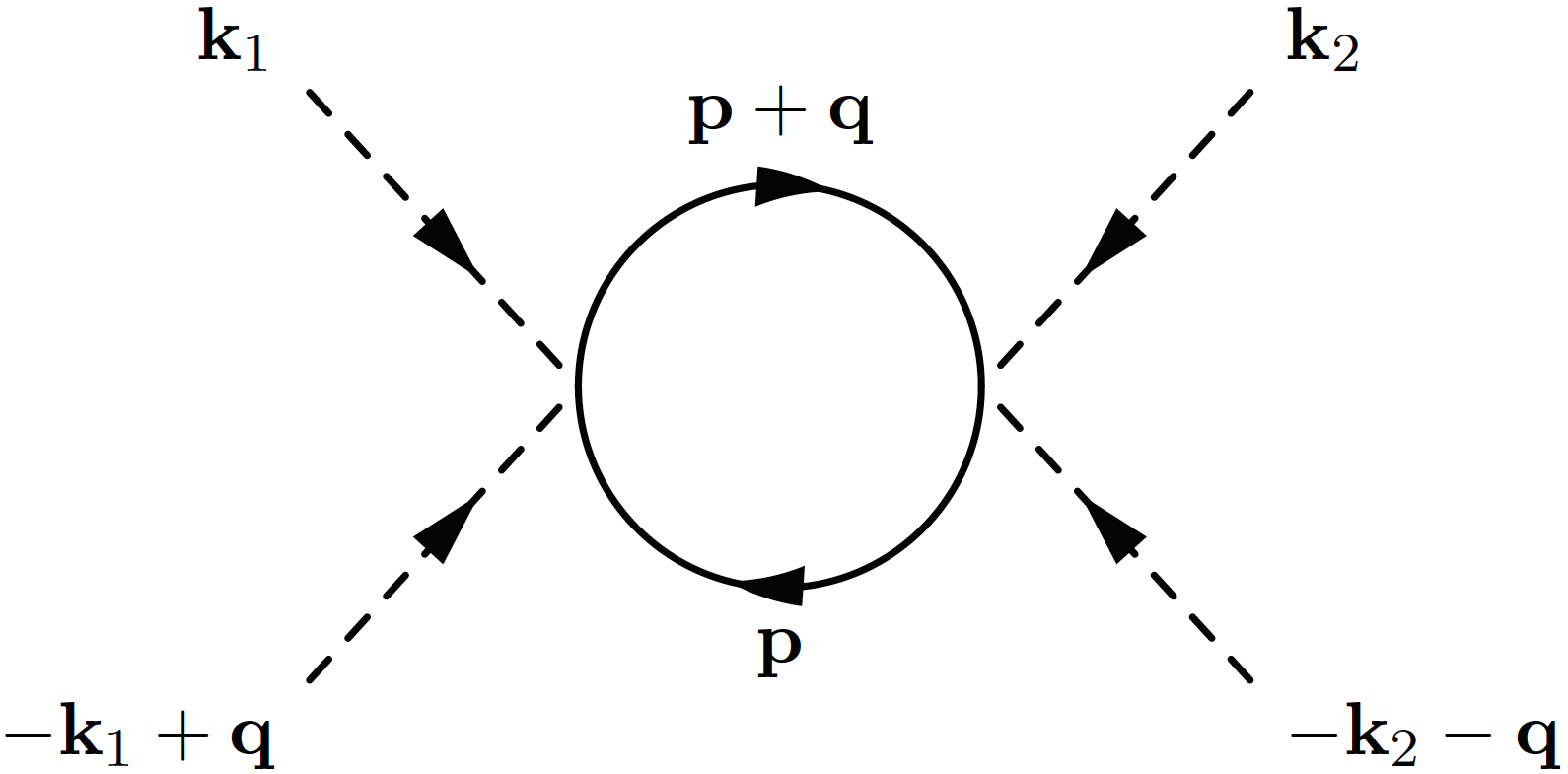}
\caption{One-loop contribution to the collapsed trispectrum of the primordial curvature perturbation. Dashed lines represent $\pi$, and solid lines represent $s$.}
\label{fig:firstfeynman}
\end{figure}

Consider a theory of inflation that consists of two fields, the inflaton $\phi$ and a massive scalar $s$. Working in the gauge where $\phi(x)=\phi_0(t)$, the Lagrangian describing the Goldstone mode $\pi$ due to the breaking of time translational invariance and $s$ can be written as
  \begin{equation}
\label{eq:int}
{\cal L} = \frac{1}{2}g^{\mu \nu}\partial_\mu \pi \partial_\nu \pi +\frac{1}{2}g^{\mu\nu}\partial_\mu s\partial_\nu s- {m^2 \over 2} s^2 + \frac{1}{\Lambda^{2}}g^{\mu\nu}\partial_{\mu}\pi\partial_{\nu}\pi s^{2} + \dots,
\end{equation}
where the action is $S = \int d^{4}x\sqrt{-g}\mathcal{L}$.  The dimension six operator in (\ref{eq:int}) induces the one-loop contribution to the four-point function of $\pi$ depicted in Fig.~\ref{fig:firstfeynman}. The complete theory includes additional interactions denoted by the ellipsis above~\cite{Senatore:2010wk,Khosravi:2012qg}\footnote{For example, the interaction $ 2\dot{\phi}_{0}\partial_{\tau}\pi s^{2}/ \Lambda^{2}$ will also appear.}, which will give rise to other one-loop contributions that are comparable to or may even dominate this diagram. The goal of this section is to illustrate the infrared behavior of loop contributions to the correlation functions of $\pi$, which have interesting implications for the correlation functions of galaxies.  For simplicity, we only consider the interaction given in (\ref{eq:int}) and leave a more complete study to future work. 

We focus on the ``collapsed'' limit of the diagram, which occurs when the external wavevectors come in pairs that are nearly equal and opposite, as shown in Fig.~\ref{fig:firstfeynman} with $q\ll k_i$.  This contribution to the four-point function has previously been computed in \cite{Arkani-Hamed:2015bza}, where the role of conformal symmetry was emphasized. In this section, we review this calculation and describe its effect on the power spectrum of galaxy overdensities.

To begin, we express the quantum fields $\pi$ and $s$ in terms of creation and annihilation operators
\begin{align}\label{eq:modesec2}
 \pi({\bf x},\tau)&=\int {d^3 k \over (2 \pi)^3}  a({\bf k})  \pi_{k}(\eta) e^{i{\bf k}\cdot {\bf x}}  +{\rm h.c.}  \,,   \quad 
 s({\bf x},\tau) =\int {d^3 k \over (2 \pi)^3}  b({\bf k})  {s}_{k}(\eta) e^{i{\bf k}\cdot {\bf x}} +{\rm h.c.}  \,,
\end{align}
where $k = |\bf{k} |$, and $\eta=k \tau$ for conformal time $\tau<0$. The mode functions satisfy the equations of motion of the free theory with appropriate boundary conditions and are
\begin{align}
\pi_k(\eta)&={H \over k^{3/2}}\pi(\eta)\,, \qquad \pi(\eta)=\frac{1}{\sqrt{2}}(1+i\eta)e^{-i \eta} \,, \\[5pt]
s_k(\eta)&={H \over k^{3/2}}s(\eta)\,, \qquad s(\eta)=-ie^{i\left(2-\nu\right)\frac{\pi}{2}}{\sqrt{\pi}\over 2}(-\eta)^{3/2}H^{(1)}_{\frac32 - \nu}(-\eta) \,,
\end{align}
where $\nu=3/2- \sqrt{9/4-m^2/H^2}$ and $H^{(1)}_z$ is the Hankel function of the first kind. We assume that the mass $m$ of the field $s$ is much less than the Hubble constant $H$ during inflation, or equivalently $\nu \ll 1$.\footnote{In (\ref{eq:int}), the mass $m$ includes contributions from terms such as $(\dot\phi_0^2/\Lambda^2)s^2$.  Tuning is required for $m\ll H$.}
We are interested in this region of parameter space because it leads to the largest infrared enhanced contributions to the four-point function. 

Let us now compute the contribution in Fig.~\ref{fig:firstfeynman} to the collapsed trispectrum of the primordial curvature perturbation $\zeta = -(H/ \dot{\phi}_0) \pi$.  The primordial curvature trispectrum $T_{\zeta}$ is defined by
\begin{align}\label{eq:tri}
\langle \zeta_{{\bf k}_1}\zeta_{{\bf k}_2}\zeta_{{\bf k}_3}\zeta_{{\bf k}_4}\rangle_c=T_{\zeta}({\bf k}_1,{\bf k}_2,{\bf k}_3,{\bf k}_4)(2\pi)^3\delta^{3}({\bf k}_1+{\bf k}_2+{\bf k}_3+{\bf k}_4) \,
\end{align}
where the subscript $c$ denotes the connected part of the four-point function.  In Fig.~\ref{fig:firstfeynman} ${\bf k}_3=-{\bf k}_1+{\bf q}$ and ${\bf k}_4=-{\bf k}_2-{\bf q}$.  The collapsed configuration $T_\zeta^{\rm coll}$ occurs when $q \ll k_{i}$.

Using the in-in formalism~\cite{Weinberg:2005vy} and introducing the variables $\eta=k_1 \tau$ and $\eta'=k_2\tau'$ we find
\begin{align}
\label{eq:bigeq}
T^{\, \rm coll}_{\zeta} \ &=32\left({H \over \Lambda}\right)^4 \left({H^2 \over \dot{\phi}_0}\right)^4  {1 \over k_1^3 k_2^3} \int {d^3 p \over (2 \pi)^3}{ 1 \over |{\bf p}+ {\bf q}|^3 p^3} \int_{-\infty}^0 {d \eta \over \eta^2}  \int_{-\infty}^{{k_1\over k_2} \eta } {d \eta' \over \eta'^2}e^{\epsilon (\eta +\eta')}{\rm Im}\left[F(\eta) \right] 
 \nonumber \\[5pt]
& \times   {\rm Im}\left[ F(\eta') s\left({|{\bf p}+ {\bf q}| \over k_1} \eta \right)s^*\left( {|{\bf p}+ {\bf q}| \over k_2} \eta' \right)s\left( { p\over k_1} \eta \right)s^*\left({p \over k_2} \eta' \right) \right]  + \big( k_1 \leftrightarrow k_2 \big)  \,
\end{align}
where
\begin{equation}
F(\eta)= \pi(0)^2 \left( [\partial_{\eta}\pi^*(\eta)]^2 -  [\pi^*(\eta)]^2 \right) \,.
\end{equation}
In Eq.~(\ref{eq:bigeq}), $\epsilon$ is an infinitesimal positive quantity that regulates the time integrations in the distant past and we have expanded in $q \ll k_{i}$.

The dominant contribution of the loop integral in (\ref{eq:bigeq}) comes from $p \sim q$.  Moreover, the time integrals are dominated at late times $\eta\,, \eta^\prime \sim - 1$. We can thus use the small $\eta$ expansion of the $s$ mode function
\begin{align}\label{eq:smodeexpand}
s(\eta)\stackrel{\eta\rightarrow0} \simeq b_1 (-\eta)^\nu \,, \qquad |b_1|^2=2^{1-2\nu}\Gamma(3/2-\nu)^2/\pi\stackrel{\nu\rightarrow0}{\simeq}1/2 \ 
\end{align}
to find
\begin{equation}
T^{\, \rm coll}_{\zeta} \simeq 8 \left({H \over \Lambda}\right)^4 \left({H^2 \over \dot{\phi}_0}\right)^4  {1 \over (k_1 k_2)^{3+2 \nu}}I_{2\nu}(q)J^2 \, 
\end{equation}
where
\begin{align}\label{eq:integrals}
I_{2\nu}(q)&= \int{d^3 p \over (2 \pi)^3}{1 \over |{\bf p}+ {\bf q}|^{3-2 \nu} p^{3-2 \nu}}\stackrel{\nu\rightarrow0}{\simeq}{1 \over 2 \pi^2} {1 \over \nu}q^{-3+4 \nu} \,, \\[5pt]
J&=\int_{-\infty}^0{d\eta \over \eta^2}e^{\epsilon \eta}(-\eta)^{2 \nu}{\rm Im}\left[      
F(\eta)
\right]= 2^{-2-2\nu}\frac{\Gamma(2+2\nu)}{1-2\nu}\stackrel{\nu\rightarrow0}{\simeq} {1 \over 4} \,.
\end{align}
In (\ref{eq:integrals}) we have kept only the term singular in $\nu$.
Note that our result is finite because we focused on the relevant region $p \sim q \ll k_{i}$ and neglected the region of large loop momenta which is not as important in the limit $q \rightarrow 0$. The UV divergence due to the region of large loop momentum would be rendered finite by a counterterm.

Our final result for the four-point function of the curvature perturbation for $m \ll H$ and $q \ll k_{i}$ is 
\begin{equation}
\label{result1}
T_{\zeta}^{\, \rm coll} \simeq {1 \over 4 \pi^2}{1 \over \nu}\left({H \over \Lambda} \right)^4\left({H^2 \over \dot \phi_0}\right)^4{1 \over k_1^3 k_2^3 q^3}\left({q^2 \over k_1 k_2} \right)^{2 \nu} \,.
\end{equation}
The factors of wavevector magnitudes in (\ref{result1}) essentially follow from the form of $s(\eta)$ expanded for small $\eta$ in the limit $m \ll H$, and from dimensional analysis. For $m \ll H$ the four-point function is enhanced by $1/\nu \simeq 3 H^2/m^2$. This arises because for small $m/H$ the the mode function $s(\eta)$ falls off slowly as the mode $k$ redshifts outside the de-Sitter horizon. Note also that there is no IR divergence in the loop integration since the $s$ field is massive. Three- and four-point curvature fluctuations generated by loop effects have been considered in Refs.~\cite{Cogollo:2008bi,Rodriguez:2008hy,Kumar:2009ge,Bramante:2011zr} using the $\delta N$ formalism. It would be interesting to see if this method  can reproduce (\ref{result1}). 

We now qualitatively discuss the effects of (\ref{result1}) on the galaxy power spectrum.  To begin, the matter overdensity $\delta_R$ averaged over a spherical volume of radius $R$ is related to the primordial curvature fluctuation via
\begin{align}
\label{delta to xi main text}
\delta_{R}({\bf k}) 
=\frac{2k^2}{5\Omega_m H_0^2}T(k)W_R(k)\zeta_{{\bf k}}
\end{align}
where $W_R(k)$ is the window function, $T(k)$ is the transfer function, $\Omega_{m}$ is the ratio of the matter density to the critical density today, and $H_{0}$ is the Hubble constant evaluated today.    

We consider an expansion for the galaxy overdensity $\delta_h$ in terms of $\delta_R$ of the following form
\begin{equation}
\label{bias expansion}
\delta_{h}({\bf x}) = b_{1} \delta_{R} ({\bf x}) + b_{2} (\delta_{R}^2({\bf x})-\sigma_R^2 )  + b_3( \delta_R^3({\bf x}) -  3\delta_R ({\bf x})\sigma_R^2 ) +   \dots \,,
\end{equation}
where $\sigma_R^2=\langle \delta_R({\bf x})\delta_R({\bf x})\rangle$ and the constants $b_{1}$, $b_{2}$, and $b_3$ are bias coefficients (for a more complete treatment, see \cite{Desjacques:2016bnm}).  The bias coefficients can be determined from data or computed using a specific model of galaxy halo formation that expresses the galaxy overdensity in terms of  $\delta_R$. The two-point function of the galaxy overdensity is then:
\begin{align}
\label{2 point bias expan}
\left<\delta_{h}({\bf x})\delta_{h}({\bf y})\right> &=b_{1}^{2}\left<\delta_{R}({\bf x})\delta_{R}({\bf y})\right>   
+b_{1}b_{2} \big( \left< (\delta_{R}^{2}({\bf x})- \sigma_R^2)\delta_{R}({\bf y})\right> +  \left<\delta_{R}({\bf x})(\delta_{R}^{2}({\bf y})-\sigma_R^2)\right> \big) \nonumber\\
&+b_{2}^{2}\left<(\delta_{R}^{2}({\bf x})-\sigma_R^2)(\delta_{R}^{2}({\bf y})-\sigma_R^2)\right> + \dots 
\end{align}
A similar expression could be derived for the galaxy-matter cross-correlation $\langle \delta_h({\bf x})\delta_R({\bf y}) \rangle$.  

Ignoring other contributions to the non-Gaussianities of $\zeta$ besides the one given in (\ref{result1}), the term proportional to $b_2^2$ in (\ref{2 point bias expan}) yields a contribution to the galaxy power spectrum of the form $P_{hh}(q) \sim 1/q^{3-4\nu}$, but not to the galaxy-matter cross-correlation $P_{hm}(q)$. Hence this loop contributes to stochastic bias, but not to scale-dependent bias. Note that in the absence of primordial non-Gaussianity, $P_{hh}(q)\sim q$, so the trispectrum contribution is enhanced by a relative factor of $q^{-4+4\nu}$ and dominates as $q \rightarrow 0$.

It is worth emphasizing that we have only considered one particular interaction in this theory, and have ignored other interactions which may give even more important contributions to stochastic and scale-dependent bias.  We now turn to a model within QSFI in order to make a full prediction in a consistent theory.

\section{Loop-Induced Stochastic Bias in Quasi-Single Field Inflation}
\label{qsfi section}

In this section, we show that loop-induced non-Gaussianities in QSFI~\cite{Chen:2009zp} can give rise to stochastic bias that is potentially observable given the stringent constraints from CMB data on non-Gaussianities.  The model we consider consists of an inflaton $\phi$ and a massive scalar $s$ with the symmetries $\phi \to \phi + c$, $\phi \to -\phi$, and $s \to -s$. These symmetries are broken by the potential of $\phi$ as well as by the lowest dimension operator that couples $\phi$ and $s$, $g^{\mu \nu}\partial_\mu \phi \partial_\nu \phi s/\Lambda$.  The Lagrangian written in terms of the Goldstone mode $\pi$ is
 \begin{equation}\label{qsfi Lagrangian}
{\cal L}=\frac{1}{2}g^{\mu \nu}\partial_\mu \pi \partial_\nu \pi \left(1+\frac{2}{\Lambda}s\right)+\frac{1}{2}g^{\mu\nu}\partial_\mu s\partial_\nu s-\mu H \tau s\partial_\tau \pi- {m^2 \over 2} s^2 - {V^{(4)}  \over 4!} s^4  \,
\end{equation}
where the kinetic mixing term is parameterized by the coupling $\mu=2\dot\phi_0/\Lambda$ and we have ignored higher order terms in the potential for $s$. Similar to the previous section, we focus here on the region where $m \ll H$ and $\mu \ll H$, which gives the most significant long wavelength enhancement to the galaxy power spectrum.

\begin{figure}
\includegraphics[width=4in]{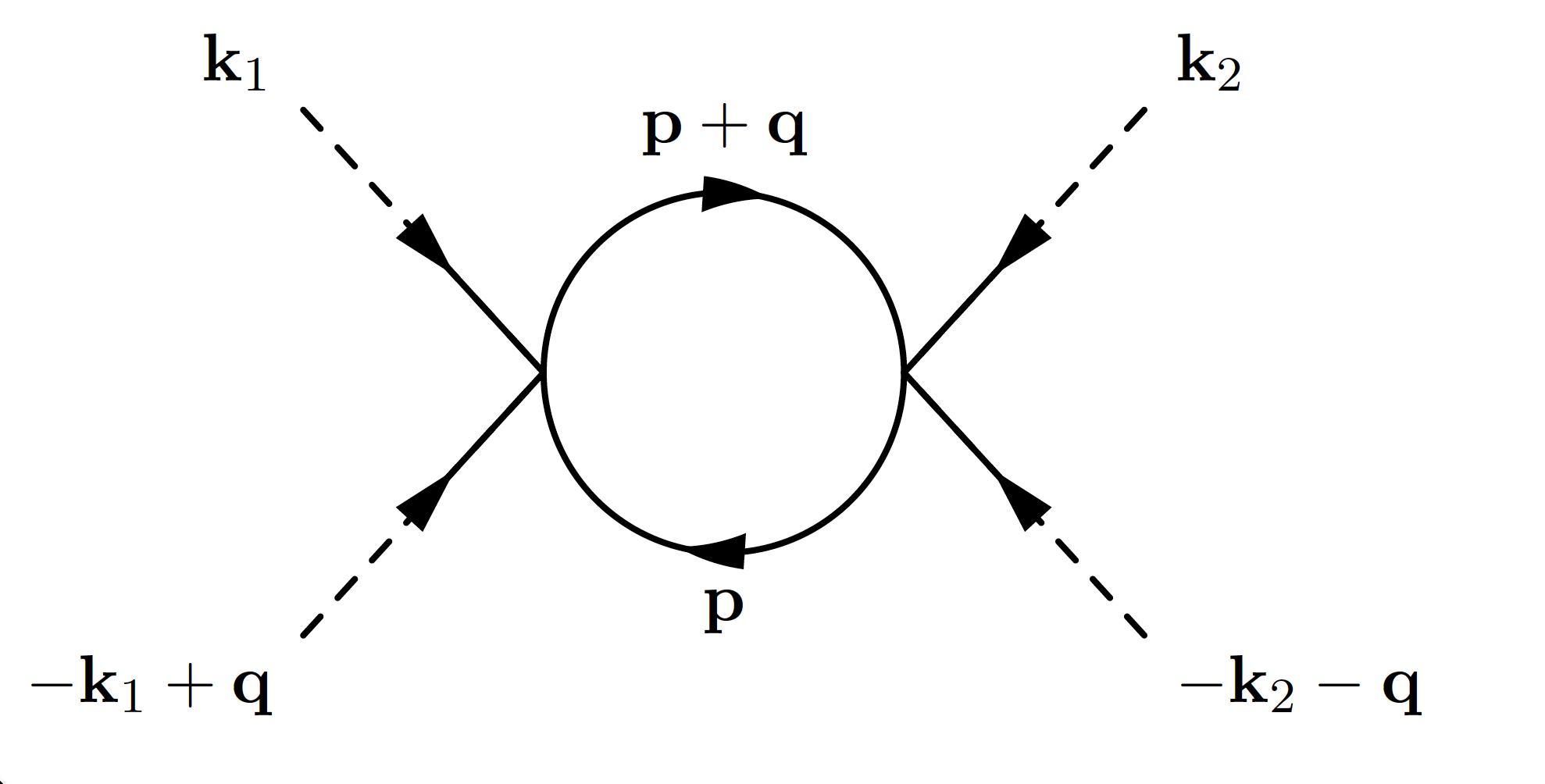}
\caption{One-loop contribution to the collapsed trispectrum of the primordial curvature perturbation in QSFI. Dashed lines represent $\pi$, and solid lines represent $s$.}
\label{feynmanloop}
\end{figure}

Due to the kinetic mixing, $\pi$ and $s$ share a set of creation and annihilation operators:
\begin{align}\label{eq:mode}
 \pi({\bf x},\tau)&=\int {d^3 k \over (2 \pi)^3} \left( a^{(1)}({\bf k})  \pi_{k}^{(1)}(\eta) e^{i{\bf k}\cdot {\bf x}}+a^{(2)}({\bf k}) \pi_{k}^{(2)}(\eta)e^{i{\bf k} \cdot {\bf x}}  +{\rm h.c.}  \right) \, \\
 s({\bf x},\tau)&=\int {d^3 k \over (2 \pi)^3} \left( a^{(1)}({\bf k})  {s}_{k}^{(1)}(\eta) e^{i{\bf k}\cdot {\bf x}}+a^{(2)}({\bf k}) {s}_{k}^{(2)}(\eta)e^{i{\bf k} \cdot {\bf x}}  +{\rm h.c.}  \right) \, .
 \end{align}
The mode functions $\pi_k^{(i)}= (H/k^{3/2})\pi^{(i)}$ and $s_k^{(i)} =  (H/k^{3/2})s^{(i)}$ are difficult to solve for exactly.  However, analytic progress can be made by considering series solutions.  It can easily be checked that the most general series solutions to the mode equations derived from (\ref{qsfi Lagrangian}) are
\begin{align}\label{small mode function behavior}
\pi^{(i)}(\eta) &= \sum_{n=0}^\infty \left[  a_{0,2n}^{(i)}(-\eta)^{2n} + a_{-,2n}^{(i)}(-\eta)^{2n+\alpha_-} +  a_{+,2n}^{(i)}(-\eta)^{2n+\alpha_+} +  a_{3,2n}^{(i)}(-\eta)^{2n+3} \right] \, \\ \label{small s behavior}
s^{(i)}(\eta) &= \sum_{n=0}^\infty \left[  b_{0,2n}^{(i)}(-\eta)^{2n} + b_{-,2n}^{(i)}(-\eta)^{2n+\alpha_-} +  b_{+,2n}^{(i)}(-\eta)^{2n+\alpha_+} +  b_{3,2n}^{(i)}(-\eta)^{2n+3} \right] \,
\end{align}
where 
$\alpha_\pm=3/2\pm\sqrt{9/4-\mu^2/H^2-m^2/H^2}$ and $b_{0,0}^{(i)} = 0$.  For ease of notation we denote $a^{(i)}_{r,0}$ and $b^{(i)}_{r,0}$ as $a^{(i)}_r$ and $b^{(i)}_r$. In Ref. \cite{An:2017rwo}, it was shown that the non-Gaussianities can be well approximated by a finite set of combinations of the power series coefficients when $\mu,m \ll H$.
The combinations of power series coefficients needed to compute the loop in Fig.~\ref{feynmanloop} are
\begin{align}\label{IR relations}
 {\rm Re}\left[a_0^{(i)}b_-^{*(i)}\right] &\simeq \frac{-3\mu H}{2(\mu^2+m^2)} \,, \quad {\rm Im}\left[a_0^{(i)}b_3^{*(i)}\right]= \frac{\mu H}{2(\mu^2+m^2)} \,, \quad
\big| b^{(i)}_- \big|^2\simeq \frac{1}{2} \,, \\[5pt]
 {\rm Im}\left[a_0^{(i)}b_-^{*(i)}\right] &= {\rm Im}\left[a_0^{(i)}b_{0,2}^{*(i)}\right]={\rm Im}\left[a_0^{(i)}b_{-,2}^{*(i)}\right]={\rm Im}\left[a_0^{(i)}b_{+}^{*(i)}\right]=0 \,,
\end{align}
which were determined in \cite{An:2017rwo}. The repeated superscripts $(i)$ are summed over $i=1,2$.  The above expressions are valid for $\mu/H$, $m/H\ll1$.

We can now compute the loop contribution to the collapsed limit of the curvature perturbation trispectrum shown in Fig.~\ref{feynmanloop}. Again, using the in-in formalism and the variables $\eta=k_1 \tau$ and $\eta'=k_2\tau'$, we find
\begin{align}
T_\zeta^{\rm coll}
&=2{V^{(4)}}^2 \left(\frac{H^2}{\dot \phi_0}\right)^4\frac{1}{k_1^3 k_2^3} \int \frac{d^3 p}{(2\pi)^3}\frac{1}{|{\bf p}+{\bf q}|^3 p^3} \int_{-\infty}^0\frac{d\eta}{\eta^4}\int_{-\infty}^{{k_1 \over k_2}\eta} \frac{d\eta'}{\eta'^4}{\rm Im}\left[(\pi^{(i)}(0)s^{*(i)}(\eta))^2\right] \nonumber \\[5pt]
&\quad \times 
{\rm Im}\left[[\pi^{(j)}(0)s^{*(j)}(\eta')]^2   s^{(k)}\left(\frac{|{\bf p}+{\bf q}|}{k_1}\eta\right)s^{*(k)}\left(\frac{|{\bf p}+{\bf q}|}{k_2}\eta' \right)s^{(l)}\left( { p \over k_1} \eta \right)s^{*(l)}\left( {p \over k_2} \eta' \right) \right]\nonumber \\[5pt]
&\quad + \big( k_1 \leftrightarrow k_2 \big) \, .
\end{align}
Similar to before, the dominant contribution to the loop integral occurs for loop momenta $p \sim q \ll k_{i}$ and  the time integrals are dominated by late times. We can immediately expand the $s$ mode functions to find
\begin{align}
&T_\zeta^{\rm coll} \simeq \frac12 {V^{(4)}}^2\left(\frac{H^2}{\dot \phi_0}\right)^4\frac{1}{(k_1 k_2)^{3+2\alpha_-}}I_{2\alpha_-}(q)K(\mu,m)^2 \,,
\end{align}
where $I_\nu(q)$ is given in (\ref{eq:integrals}) and
\begin{equation}\label{qsfi time integral}
K(\mu,m)=\int_{-\infty}^0d\eta (-\eta)^{-4+2\alpha_-}{\rm Im}\left[(\pi^{(i_1)}(0)s^{*(i_1)}(\eta))^2\right] \,.
\end{equation}
It was shown in~\cite{An:2017rwo} that the most important contribution to (\ref{qsfi time integral}) is obtained by cutting off the lower bound of the integral at $\eta_{0}$ which is around horizon crossing. Inserting the power series expansions of the mode functions in (\ref{small mode function behavior}) and (\ref{small s behavior}), we find
\begin{equation}\label{eq:Kint}
K(\mu,m)\simeq2 \ {\rm Im}\left[a_0^{(i)}b_3^{*(i)}\right]{\rm Re}\left[a_0^{(j)}b_-^{*(j)}\right]\int_{\eta_{0}}^0  d\eta (-\eta)^{-1+3\alpha_-}  \simeq-\frac{2}{3}\frac{(3\mu/2)^2H^4}{(\mu^2+m^2)^3} \,,
\end{equation}
where we have neglected contributions from higher powers of $\eta$ which are suppressed in the limit $\alpha_- \ll 1$.
Note that this piece most singular in $\alpha_-$ is insensitive to the choice of $\eta_{0}$. 
Our final result for the four-point function of the curvature perturbation for $m\,, \mu \ll H$ and $q \ll k_{i}$ is then
\begin{align}\label{eq:qsfi}
&T_\zeta^{\rm coll} \simeq \frac{1}{3\pi^2}{V^{(4)}}^2\left(\frac{H^2}{\dot \phi_0}\right)^4\frac{1}{k_1^3 k_2^3 q^{3}}\left(\frac{q^2}{k_1 k_2}\right)^{2\alpha_-}\frac{(3\mu/2)^4H^{10}}{(\mu^2+m^2)^7} \, .
\end{align}
In (\ref{eq:qsfi}), the factors of wavevector magnitudes and $\alpha_-^{-1}$ from the integral $I_{2\alpha_-}$ are the same as those in (\ref{result1}) from the integral $I_{2\nu}$. These features are characteristic of quantum mechanical effects from the exchange of a massive particle~\cite{Arkani-Hamed:2015bza,Mirbabayi:2015hva}.

We now consider the long wavelength enhancement to the galaxy power spectrum resulting from this collapsed primordial trispectrum.
In our numerical evaluation, we make the simplifying assumption that galaxies form at points in space at which the smoothed matter overdensity is greater than a threshold density at the time of collapse $\delta_c(a_{\rm coll})$, \textit{i.e.} $n_h({\bf x})\propto\Theta_H\left(\delta_R({\bf x}, a_{\rm coll})-\delta_c(a_{\rm coll})\right)=\Theta_H\left(\delta_R({\bf x})-\delta_c\right)$, where $\delta_c\equiv\delta_c(a_{\rm coll})/D(a_{\rm coll})$.\footnote{$\delta_R({\bf x})$ is the linearly evolved matter overdensity today.}  We further assume that  $\delta_c (a_{\rm coll})=1.686$ \cite{Gunn:1972sv}, all halos collapse instantaneously at redshift $z=1.5$, and their number density does not evolve in time after collapse.  This corresponds to a value of $\delta_c=4.215$. The galaxy overdensity is defined by $\delta_h({\bf x})=(n_h({\bf x})-\langle n_h\rangle)/\langle n_h \rangle$.  With this threshold collapse model, the bias coefficients are given by (see e.g.~\cite{Ferraro:2012bd})
\begin{align}
b_1=\frac{e^{-\frac{\delta_c^2}{2\sigma_R^2}}}{\sqrt{2\pi}\sigma_R \langle n_h \rangle}\,, \quad b_2=\frac{\delta_c}{\sigma_R}\frac{e^{-\frac{\delta_c^2}{2\sigma_R^2}}}{2!\sqrt{2\pi}\sigma_R^2 \langle n_h \rangle}\,, \quad 
b_3=\left(\frac{\delta_c^2}{\sigma_R^2}-1\right)\frac{e^{-\frac{\delta_c^2}{2\sigma_R^2}}}{3!\sqrt{2\pi}\sigma_R^3 \langle n_h \rangle}
\end{align}
where $\langle n_h \rangle={\rm erfc}\left(\delta_c/(\sqrt{2}\sigma_R)\right)/2$.  
We use the BBKS approximation to the transfer function \cite{Bardeen:1985tr} and the top-hat window function $W_R(k)=3(\sin(kR)-kR \cos(kR))/(kR)^3$. Moreover, we take $R=1.9\ {\rm Mpc}/h$ as the smoothing scale, and numerically we find $\sigma_R=3.62$.

The Fourier transform of $\langle \delta_R({\bf x})\delta_R({\bf y})\rangle$ gives the matter power spectrum $P_{mm}(q)$:
\begin{align}
P_{mm}(q)=\left(\frac{2}{5\Omega_m H_0^2 }\right)^2\left(\frac{H^2}{\dot \phi_0}\right)^2C_2(\mu,m)T(q)^2 q \,,
\end{align}
where $C_2(\mu,m)=1/2+2{(3\mu/2)^2H^2}/{(\mu^2+m^2)^2}$~\cite{An:2017rwo}.
It then follows from (\ref{2 point bias expan}) that the ratio of the galaxy power spectrum to the matter power spectrum normalized by $b_1^2$ is
\begin{align}\label{PhhPmm}
{ P_{hh}(q) \over b_1^2P_{mm}(q)} = 1+\frac{b_2^2}{b_1^2}\left(\frac{2}{5 \Omega_m H_0^2 R^2}\right)^2\left(
\frac{H^2}{\dot \phi_0}\right)^2\frac{{V^{(4)}}^2 {\cal J}^2}{3\pi^2}\frac{(q R)^{-4+4\alpha_-}}{T(q)^2 }\frac{(3\mu/2)^4H^{10}}{(\mu^2+m^2)^7C_2(\mu,m)}
\end{align}
where
\begin{align}
\label{jequation}
{\cal J}&=\frac{1}{2\pi^2}\int_0^\infty du \ T\left( u / R \right)^2W_R\left( u / R \right)^2 u^3.
\end{align}

\begin{figure}
\includegraphics[width=6in]{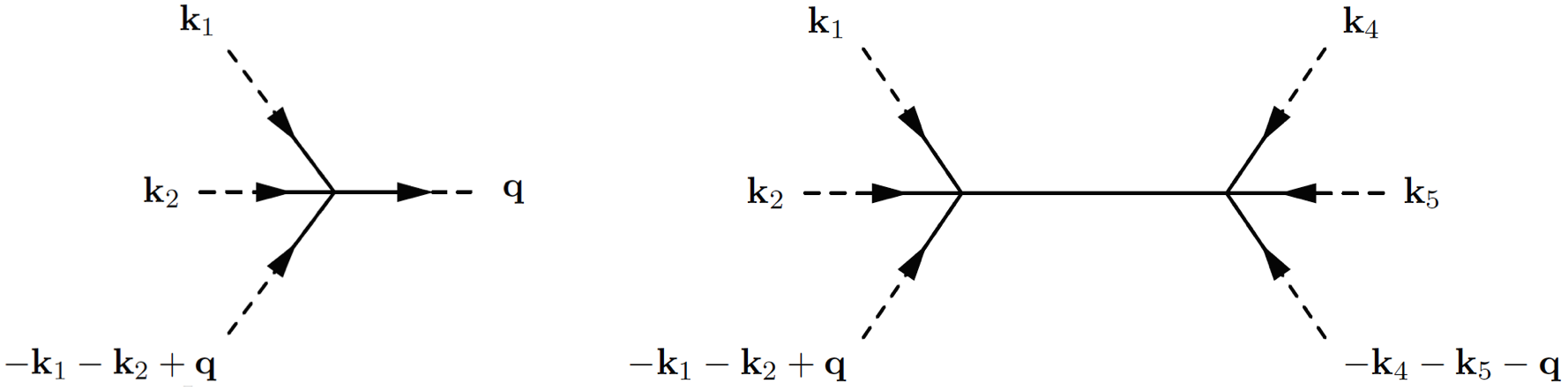}
\caption{These two tree-level diagrams involving the $V^{(4)}$ interaction can also contribute to scale-dependent and stochastic bias. However, these contributions are small compared to the loop contribution in Fig.~(\ref{feynmanloop}) due a suppression arising from the integration over additional hard external wavevectors.}
\label{feynmanother}
\end{figure}

\begin{figure}
\includegraphics[width=4in]{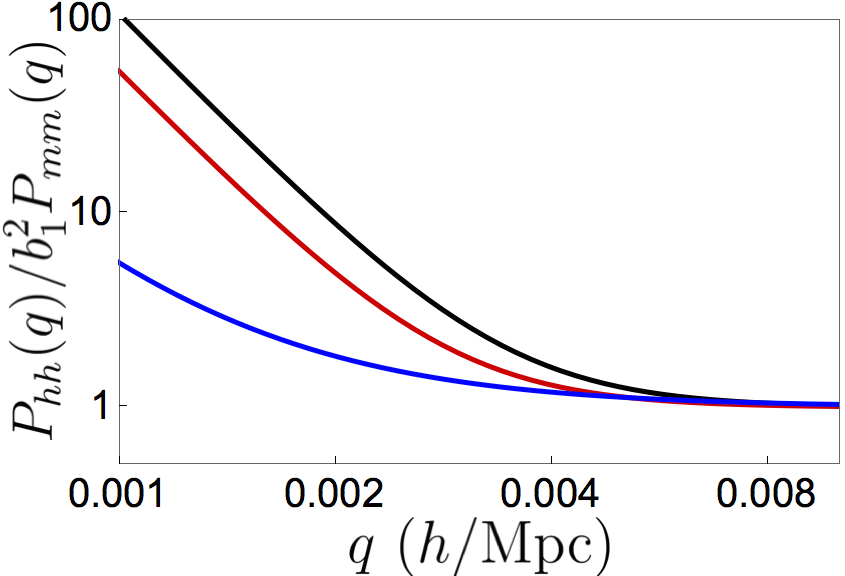}
\caption{The ratio $P_{hh}(q)/b_1^2 P_{mm}(q)$ is plotted for $\tau_{NL}^{2\sigma}=2800$ (Planck 2013) in black, and $\tau_{NL}^{2\sigma}/2=1400$ in red.  In blue, we plot the power spectrum ignoring the loop contribution and considering only the tree diagrams in Fig. \ref{feynmanother}, using the $\tau_{NL}^{2\sigma}$ bound.  Note that the enhanced behavior begins around $(200 \ {\rm Mpc}/h)^{-1}$ for the black curve, and around $(300 \ {\rm Mpc}/h)^{-1}$ for the red curve. Moreover, note that the tree contributions in blue are very small compared to the loop contribution in black.  We plot for $\mu/H=m/H=0.274$, corresponding to $\alpha_-=0.05$.  Moreover we take $R=1.9 \ {\rm Mpc}/h$ and $\delta_c=4.215$.}
\label{Phh}
\end{figure}

The $V^{(4)}$ interaction in (\ref{qsfi Lagrangian}) also gives rise to the tree-level diagrams shown in Fig.~\ref{feynmanother} which contribute to the long wavelength enhancement of the galaxy power spectrum.  However, these terms contain integrals with three transfer functions rather than two like in (\ref{jequation}).  This integral then gives $\sim{\cal J}^{3/2}$ rather than ${\cal J}$. Numerically we find ${\cal J}\approx3.1\times 10^{-5}$ so the contributions from these tree-level diagrams are suppressed, as can be seen in Fig.~\ref{Phh}.

One could also consider the contribution of the $ (\partial \pi)^2s/\Lambda$ interaction in (\ref{qsfi Lagrangian}) to $P_{hh}(q)$.  However, estimating $f_{NL}=5 B_\zeta(k,k,k)/18 P_\zeta(k)^2$ from this interaction numerically, we find that $f_{NL}\lesssim 10^{-2}$ for $\mu/H$, $m/H\lesssim 0.4$.  This small $f_{NL}$ has a negligible contribution to $P_{hh}(q)$ compared to the loop contribution we have considered.

We can constrain $V^{(4)}$ using the bounds on $\tau_{NL}$ and $g_{NL}$ from Planck 2013 and 2015~\cite{Planck 2013,Planck 2015}. The bound due to $\tau_{NL}$ is estimated using (\ref{eq:qsfi}), with factors of $(q/k)^{\alpha_-}$ set to 1 in order to match the $\tau_{NL}$ shape. The bound due to $g_{NL}$ is estimated using the tree-level four-point diagram with a single $V^{(4)}$ vertex, with factors of $(k_i/k_j)^{\alpha_-}$ set to 1 to match the $g_{NL}$ shape. We take $\tau_{NL}^{2\sigma}=2.8 \times 10^{3}$ and $g_{NL}^{2\sigma}=-2.44 \times 10^{5}$ as the maximum allowed values of $\tau_{NL}$ and $g_{NL}$ at a $2\sigma$ confidence level. We find that for most of the $(\mu,m)$ parameter space $\tau_{NL}^{2\sigma}$ gives the stronger constraints on $V^{(4)}$. For $\mu/H=m/H=0.274$ (so that $\alpha_-=0.05$), we find that the $\tau_{NL}^{2\sigma}$ constraint yields $V^{(4)}\le 0.014$. 

In Fig.~\ref{Phh}, we plot the ratio $P_{hh}(q)/b_1^2 P_{mm}(q)$. The enhanced behavior begins at around $q \sim (200 \ {\rm Mpc}/h)^{-1}$ and $q \sim (300 \ {\rm Mpc}/h)^{-1}$ for the values of $V^{(4)}$ that saturate the $\tau^{2\sigma}_{NL}$ (black curve) and $\tau^{2\sigma}_{NL}/2$ (red curve) bounds.  Moreover, the blue curve is the contribution due solely to the tree-level diagrams in Fig.~(\ref{feynmanother}) using the $\tau_{NL}^{2\sigma}$ bound, and is significantly smaller than the loop contribution shown in black. 

Finally we briefly comment on how our results depend on the parameters $R$ and $\delta_{c}$.  The loop contribution to $P_{hh}(q)/b_1^2P_{mm}(q)$ is insensitive to the choice of smoothing radius $R$.  The tree-level contributions in Fig. \ref{feynmanother} increase as $R$ increases, yet even for $R=2.7 \ {\rm Mpc}/h$, we find that the loop contribution remains an order of magnitude larger than the tree-level contributions.  Furthermore, since $b_2/b_1\sim \delta_c$, the second term in (\ref{PhhPmm}) goes like $\delta_c^2/ q^{4-4\alpha_-}$.  This implies that the characteristic scale $q_0$ at which the long-wavelength enhancements become significant depends on  $\delta_c$ like $q_0\sim \delta_c^{1/2}$.

\section{Concluding Remarks}
We have shown, using a particular QSFI model, that one-loop diagrams involving an intermediate light scalar can give rise to significant stochastic bias at long wavelengths. In this model, the one-loop contribution to the four-point function of primordial curvature perturbations induces a non-Gaussian contribution to the galaxy power spectrum $P_{hh}(q)$ that is five times larger than the Gaussian one at $q\sim h/(500 \ {\rm Mpc})$ for values of $\tau_{NL}$ and $g_{NL}$ at only half their current $2\sigma$ bounds. These non-Gaussianities could be observed in upcoming large-scale surveys~\cite{Abell:2009aa,Laureijs:2011gra,Dore:2014cca}.

It would be interesting to study the effects of these loop contributions to the bias within the framework of the effective field theory of inflation. At a minimum, this would require the computation of the one-loop diagram presented in section II and the ones due to the interaction $\mathcal{L}_{I} \sim \dot{\pi} s^2$.

\section*{Acknowledgements}
This work was supported by the DOE Grant DE-SC0011632 and by the Walter Burke Institute for Theoretical Physics.

\end{document}